\documentclass[11pt]{article}

\usepackage{bm}

\usepackage{amsmath,amsthm,amssymb}
\usepackage{enumerate}
\usepackage{graphicx}
\usepackage{epsfig}

\usepackage{hyperref}
\hypersetup{
    colorlinks=true,
    linkcolor=blue,
    filecolor=magenta,
    urlcolor=blue,
}

\usepackage[T2A]{fontenc}
\usepackage{inputenc}
\inputencoding{cp1251}

\usepackage{enumitem}
\usepackage{amssymb}
\usepackage{hyperref}

  \usepackage{geometry}
\newcommand*{\mailto}[1]{\href{mailto:#1}{\nolinkurl{#1}}}

\newtheorem{theorem}{Theorem}[section]
\newtheorem{definition}[theorem]{Definition}
\newtheorem{lemma}[theorem]{Lemma}
\newtheorem{example}[theorem]{Example}
\newtheorem{proposition}[theorem]{Proposition}
\newtheorem{corollary}[theorem]{Corollary}
\newtheorem{remark}[theorem]{Remark}
\newtheorem{remarks}[theorem]{Remarks}

\newcommand{\fr}{\frac}

\newcommand{\R}{{\mathbb R}}

\newcommand{\Co}{{\mathbb C}}

\newcommand{\bA}{{\bf A}}
\newcommand{\bj}{{\bf j}}
\newcommand{\cE}{{\cal E}}

\newcommand{\bE}{{\bf E}}

\newcommand{\cH}{{\cal H}}

\newcommand{\cm}{{\rm m}}

\newcommand{\al}{\alpha}

\newcommand{\om}{\omega}

\newcommand{\De}{\Delta}

\newcommand{\ext}{{\rm ext}}

\newcommand{\lam}{\lambda}
\newcommand{\Om}{\Omega}

\newcommand{\na}{\nabla}
\newcommand{\pa}{\partial}
\newcommand{\rot}{{\rm rot\5}}
\newcommand{\dv}{{\rm div\5}}
\newcommand{\const}{{\rm const}}

\newcommand{\rRe}{{\rm Re\5}}

\newcommand{\ov}{\overline}

\newcommand{\5}{{\hspace{0.5mm}}}

\newcommand{\ds}{\displaystyle}

\date{}

\numberwithin{equation}{section}

\newcommand{\ci}{\cite}
\newcommand{\la}{\label}

\newcommand{\be}{\begin{equation}}
 \newcommand{\ee}{\end{equation}}

 \newcommand{\beqn}{\begin{eqnarray}}
 \newcommand{\eeqn}{\end{eqnarray}}

\newcommand{\ba}{\begin{array}}
 \newcommand{\ea}{\end{array}}

\newcommand{\bd}{\begin{definition}}
 \newcommand{\ed}{\end{definition}}
\newcommand{\bt}{\begin{theorem}}
 \newcommand{\et}{\end{theorem}}

\newcommand{\bp}{\begin{proposition}}
 \newcommand{\ep}{\end{proposition}}

\newcommand{\bl}{\begin{lemma}}
 \newcommand{\el}{\end{lemma}}
\newcommand{\bc}{\begin{corollary}}
 \newcommand{\ec}{\end{corollary}}

\newcommand{\bex}{\begin{example}}
 \newcommand{\eex}{\end{example}}

\newcommand{\bexs}{\begin{examples}}
 \newcommand{\eexs}{\end{examples}}

\newcommand{\bexe}{\begin{exercice}}
 \newcommand{\eexe}{\end{exercice}}

\newcommand{\br}{\begin{remark} }
 \newcommand{\er}{\end{remark}}
\newcommand{\brs}{\begin{remarks}}
 \newcommand{\ers}{\end{remarks}}

\newcommand{\bce}{\begin{center}}
\newcommand{\ece}{\end{center}}

\date{}

\numberwithin{equation}{section}

\begin{document}

\
\bce
{\huge\bf On quantum jumps and attractors of
\medskip\\
the Maxwell--Schr\"odinger equations}
 \bigskip
\bigskip\bigskip

 {\Large A.I. Komech} 
\footnote { Supported by 
Austrian Science Fund (FWF): P28152-N35.}
 \smallskip
 \\
{
 \centerline {Faculty of Mathematics, University of Vienna}
  \centerline {Institute for Information Transmission Problems of  RAS}
     \centerline {Department  of Mechanics and Mathematics of Moscow State University}
 }
  \centerline {alexander.komech@gmail.com}
\ece

\bigskip
\centerline{\it To Sasha Shnirelman on occasion of his 75-th anniversary}
\bigskip
   \begin{abstract}
Our goal is the discussion of the problem of 
 mathematical interpretation of basic postulates (or `principles') of Quantum Mechanics: transitions to quantum stationary orbits, the wave-particle duality, and the probabilistic interpretation,
in the context of semiclassical self-consistent Maxwell--Schr\"odinger equations.
We 
discuss possible relations of these postulates to
 the theory of attractors of Hamiltonian nonlinear PDEs
 relying on 
a  new general  {\it mathematical conjecture} on global attractors of 
G-invariant nonlinear Hamiltonian partial differential equations with a Lie symmetry group G.

This conjecture is inspired by our results 
on global attractors of nonlinear Hamiltonian PDEs
obtained since 1990 for a list of model equations with three basic symmetry groups: the trivial group, the group of translations, and the unitary group U(1). We present sketchy these results.

\end{abstract}
  
  

\tableofcontents

\bigskip
\bigskip


\section {Quantum postulates   and  Maxwell--Schr\"odinger equations}
 The present paper is inspired by the problem of a mathematical description
of `quantum jumps', i.e., of transitions between quantum stationary
orbits.
We discuss  the problem of mathematical description of  basic postulates of
quantum theory:
\smallskip
 

A. Transitions between quantum  stationary orbits (N. Bohr, 1913).
 \smallskip
 

B. Wave-particle duality (L. de Broglie, 1923).
 \smallskip

C. Probabilistic interpretation (M. Born, 1927).
 \smallskip\\
The problem concerns the validity of these postulates
in Schr\"odinger's quantum mechanics,
  and it still remains an open problem.
These and other questions
have been frequently addressed in the 1920s and 1930s in hot discussions by Bohr, Schr\"odinger, Einstein and others \ci {B1949}.
However, a satisfactory solutions were not achieved, and
a rigorous dynamical description  of these postulates  is still
unknown.
This lack of theoretical clarity hinders the progress in the theory (e.g., in superconductivity and  in nuclear reactions),
and in
numerical simulation of many engineering processes (e.g., of laser radiation and quantum amplifiers) since a computer can solve dynamical equations, but cannot  take into account  postulates.
\smallskip

We 
discuss the possible relations of these postulates to
 the theory of attractors of Hamiltonian nonlinear PDEs
 in the context of semiclassical self-consistent Maxwell--Schr\"odinger equations  (MS),
 applying 
a  novel general  {\bf mathematical} conjecture  on global attractors of G-invariant nonlinear Hamiltonian partial differential equations with a Lie symmetry group G.

This conjecture was inspired by a number of results 
on global attractors of nonlinear Hamiltonian PDEs
obtained since 1990 for a list of model equations with three basic symmetry groups: the trivial group, the group of translations, and the unitary group U(1). We 
briefly present  these results.
However, for the Maxwell--Schr\"odinger equations, the justification of
global attraction  still remains an open problem.

\smallskip

Note that the second-quantized MS system 
is the main subject of 
Quantum Electrodynamics \ci{Sakurai}. 
Our specific attention to the semiclassical MS equations is due to the fact that for this system 
 an
extensive empirical material is available: on atomic spectra, electron diffraction, on crystals and their
 thermal and electric conductivity, etc. Therefore, one can try to find  a possible mathematical description of these phenomena in the framework of the MS system and try to prove it. So the MS system  serves as a testing ground for the development of the theory. 
The same questions are open in Quantum Field Theory.
 However, these questions obviously cannot be clarified
until they are understood in the simpler context of 
semiclassical theory.

\smallskip

The coupled Maxwell--Schr\"odinger equations
in the Coulomb gauge $ \dv \bA (x, t) \equiv 0 $ 
  read 
  as (cf. \ci{Sakurai,NW2007})
\be \la {MS}
\!\!\!\!\!\!
\left\{\ba{rcl}
\fr 1 {c ^ 2} \ddot \bA (x, t)\!\!\! &\!\!=\!\!&\!\!\! \De \bA (x, t) +\fr1c P\bj (x, t),
\quad \De A^0(x,t)= -\rho (x, t)
 \\
\\
 i \hbar \dot \psi (x, t)\!\!\!&\!\!=\!\!&\!\!\!
\fr 1 {2\cm} [- i \hbar \na\!-\! \ds \frac ec (\bA ( x, t)\!+\!\bA_\ext  ( x, t))]^2 \psi (x, t) \!+\! e (A ^ 0  (x, t)\!+\!A ^ 0_\ext (x, t)) \psi (x, t)
\ea\right|\, x \in \R ^ 3,
\ee
where $\bA_\ext ( x, t)$ and $A^0_\ext ( x, t)$ are  external Maxwell potentials,
$e<0$ is the electron charge and $c$ is the speed of light in a vacuum.
The coupling is completed by expressing the charge and current densities
in the wave function:
\be \la {rj}
\rho (x, t) = e | \psi (x, t) | ^ 2, \qquad \bj (x, t) = \fr e \cm \rRe \{\ov \psi (x, t) [- i \hbar \na- \fr ec (\bA  ( x, t)+\bA_\ext ( x, t))] \psi (x, t) \}.
\ee
These densities  satisfy the continuity identity
$
\dot \rho (x, t) + \dv \bj(x, t) \equiv 0.
$
The system (\ref {MS})
is formally Hamiltonian, with the
Hamiltonian functional (which is the energy up to a factor)
\be\la{enc}
\cH(\bm{\Pi},\bA,\psi,t)=
\fr12[\Vert \fr1c \bm{\Pi}\Vert^2
+\Vert \rot\bA\Vert^2]+(\psi, H(\bA,\psi,t)\psi),
\ee
where 
$\Vert\cdot\Vert$ stands for the norm  in the real Hilbert space
$L^2(\R^3)\otimes\R^3$ and 
the brackets
$(\cdot,\cdot)$ stand for the inner product in 
$L^2(\R^3)\otimes\Co$.
The Schr\"odinger  operator is defined by
$$
H(\bA,\psi,t):=
\fr 1 {2m} [- i \hbar \na- \ds \frac ec (\bA ( x)+\bA_\ext  ( x,t))]^2
+e(\fr12 A^0(x)+A^0_\ext (x,t)),
$$
where $A^0(x):=(-\De)^{-1}\rho$ with $\rho(x):=e|\psi(x)|^2$.
The system (\ref{MS})  can be written in the Hamiltonian form 
 with variational derivatives as
\be\la{MSH}
\left\{\ba{l}
\fr 1{c^2} \dot\bA(t)=D_{\bm{\Pi}}\cH(\bm{\Pi(t)},\bA(t),\psi(t),t), 
\,\qquad
\fr 1{c^2} \dot{\bm{\Pi}}(t)=-D_\bA\cH(\bm{\Pi}(t),\bA(t),\psi(t),t)
\medskip\\
 i\hbar\dot\psi(t)=\fr12 D_\psi\cH(\bm{\Pi}(t),\bA(t),\psi(t),t)=H(\bA(t),\psi(t),t)\psi(t),
\ea\right|,
\ee
taking into account that $(\psi, eA^0\psi)=(A^0,\rho)=
( (-\De)^{-1}\rho,\rho)$, and hence, 
$ D_\psi( \psi, eA^0\psi)=4eA^0\psi$.
Therefore, the energy is conserved in the case
of static external  potentials
\be\la{stat}
 \bA_\ext(x,t)\equiv\bA_\ext(x),\qquad
 A^0_\ext(x,t)\equiv A^0_\ext(x).
 \ee
 For instance, in the case of an atom,  $A^0_\ext(x)$ is the Coulomb potential of the nucleus, while
 $\bA_\ext(x)$ is  the vector  potential of the   nucleus magnetic field.
       The Hamiltonian (\ref {enc}) is invariant with respect to the action of the group $ U (1) $,
\be \la {U1-2}
(\bA (x), \bm {\Pi} (x), \psi (x)) \mapsto (\bA (x), \bm {\Pi} (x), \psi (x) e ^ { i \theta}), \qquad \theta \in (0,2 \pi).
\ee
\br
{\rm
The
existence of global solutions to the
Cauchy problems for
systems (\ref {MS}) in the entire space $ \R ^ 3 $ without external potentials was proved in \ci {GNS1995}
for all finite energy initial states  (\ref {enc}).
The uniqueness of  solutions 
in the energy space was proved in \ci{BT2009}.
}
\er

   \br 
   {\rm 
   The system (\ref{MS})
was essentially introduced  by Schr\"odinger   in his first articles \ci{Schr1926},
and it
underlies the entire theory of laser radiation
\ci{SZ1997}.
}
\er
 
\section{Quantum jumps   and attractors}
   In 1913, Bohr formulated the following two fundamental postulates of the quantum theory of atoms:
\smallskip \\
{\bf I.} An atom always lives in one of the quantum stationary orbits, and
sometimes it jumps from one stationary state to another:
in the Dirac notation 
\be \la {B1}
| E_n \rangle \mapsto | E_ {n '} \rangle.
\ee
{\bf II.} The atom does not radiate in  stationary orbits. Every  jump
is followed by a radiation of an electromagnetic wave with the
frequency 
\be \la {B21}
 \om_ {nn '} = \fr {E_ {n'} - E_n} \hbar = \om_ {n '} - \om_n, \qquad \om_n: = E_n / \hbar,
\ee

Both these postulates were inspired i) by stability of atoms and ii) by the Rydberg--Ritz {\it Combination Principle}. 
With the discovery of the Schr\"odinger theory in 1926,
the question aroses about the implementation of the above Bohr axioms in the new theory.
   
\subsection{Schr\"odinger theory of stationary orbits}
Besides the equation for the wave function, the Schr\"odinger theory contains a highly
nontrivial definition of stationary orbits 
(or {\it quantum stationary states})
in the case when
 {\bf the Maxwell external potentials do not depend on time}:
In this case, the Schr\"odinger equation\index{Schr\"odinger equation} 
reads as
\be \label {Sc0}
i \hbar \dot \psi (x, t) =
H \psi (t): =
\fr 1 {2 \cm} [- i \hbar \na- \ds \frac ec \bA_\ext (x)] ^ 2 \psi (x, t) + eA_\ext^0 (x) \psi (x, t).
\ee
The corresponding 
stationary orbits are defined as finite-energy solutions of the form
\be \la {SSO}
\psi (x, t) \equiv \psi (x) e ^ {- i \om t}, \qquad \om \in \R.
\ee
Substitution 
into the Schr\"odinger equation (\ref{Sc0}) leads to the famous eigenvalue problem.
\smallskip

Such definition is rather natural, since then $ | \psi (x, t) | $ does not depend on time. Most likely, this definition was suggested
by the de Broglie wave function
 for {\it free particles} $ \psi (x, t) = Ce ^ {i (kx- \om t)} $, which factorises as $ Ce ^ {ikx} e ^ {- i \om t} $. Indeed, in the case of  {\it bound particles},
it is natural to change the spatial factor $ Ce ^ {ikx} $, since the spatial properties have changed and ceased to be homogeneous.
On the other hand, the homogeneous time factor $ e ^ {- i \om t} $ must be preserved, since the external potentials are independent of time.
However, these `algebraic' arguments do not 
withdraw the question  on agreement of the Schr\"odinger definition
with the  Bohr postulate (\ref {B1})!
\smallskip

Thus, the problem on the mathematical interpretation of the Bohr postulate (\ref {B1}) in the Schr\"odinger theory  arises. 
   One of the simplest interpretation of the jump (\ref{B1})
is the long-time  asymptotics 
\be \la {BS}
\psi (x, t) \sim \psi_ \pm (x) e ^ {- i \om_ \pm t}, \qquad t \to \pm \infty,
\ee
{\it for each finite energy solution},
 where
 $ \om _- = \om_n $ and $ \om _ + = \om_ {n '} $. However, 
{\it for  the linear Schr\"odinger equation} (\ref{Sc0}),
such asymptotics are obviously  wrong, due to the {\it superposition principle}:
for example, for solutions of the form $ \psi (x, t) \equiv \psi_1 (x) e ^ {- i \om_1 t} + \psi_2 (x) e ^ {- i \om_2 t} $ with $\om_1\ne\om_2$.
It is exactly this contradiction which shows that the linear 
Schr\"odinger equation alone cannot  serve as a basis for the 
theory compatible
with the Bohr postulates.
Our main conjecture is that these asymptotics are inherent properties of the nonlinear Maxwell--Schr\"odinger equations (\ref{MS}). This conjecture
is suggested by the following perturbative arguments.

\subsection{Bohr's postulates by perturbation theory}   
   
The remarkable success of the Schr\"odinger theory was in the explanation
of  Bohr's postulates via asymptotics (\ref{BS})
 by 
 means of 
{\it perturbation theory} applied to the 
{\it coupled Maxwell--Scr\"odinger equations} (\ref {MS})
in the case of {\it static external potentials}.
Namely, as a first approximation, the fields $ \bA (x, t) $ and $ A^0 (x, t) $ in the Schr\"odinger equation of the system
 (\ref {MS})  can be neglected, so we obtain 
the equation (\ref{Sc0}).
For 
`sufficiently good'
external potentials and initial conditions,
each finite energy solution
can be expanded in  eigenfunctions
 \be \la {Sexp}
 \psi (x, t) = \sum_n C_n \psi_n (x) e ^ {- i \om_n t} + \psi_c (x, t), \qquad
 \psi_c (x, t) =
 \int C (\om) e ^ {- i \om t} d \om,
\ee
where integration is performed over the continuous spectrum of the Schr\"odinger operator $ H $, and 
 for any $R>0$ we have
\be\label{psic}
\int_{|x|<R}|\psi_c (x, t)dx\to 0,\quad t\to\pm\infty,
\ee
see, for example, \ci [Theorem 21.1] {KK2012}.
The
substitution of this expansion into the expression for currents (\ref {rj}) gives 
 \be \la {jexp}
j (x, t) = \sum_ {nn '} j_ {nn'} (x) e ^ {- i \om_ {nn '} t} + c.c. + j_c (x, t),
\ee
where $ j_c (x, t) $ has a continuous frequency spectrum. 
Thus, the currents on the right hand side of 
the
Maxwell equation from (\ref {MS}) contains,
besides the continuous spectrum, only
{\it discrete frequencies} $ \om_ {nn '} $.
Hence, the {\it discrete spectrum} of the corresponding 
Maxwell field also contains only these frequencies $ \om_ {nn '} $.
This jusitfies the Bohr rule (\ref {B21})
{\it in the first order of perturbation theory}, since this calculation ignores the back effect of radiation into the atom.
\smallskip

Moreover, these arguments also suggest to treat
 the jumps (\ref {B1}) as
the {\it single-frequency asymptotics} (\ref {BS})
for solutions of the Schr\"odinger equation 
{\it coupled to the
Maxwell equations}.

Indeed, the currents (\ref{jexp}) on the right of
the Maxwell equation from (\ref {MS}) produce the radiation when
non-zero frequencies $ \om_ {nn '} $ are present. 
This is due to the fact that $\R\setminus 0$ is the
absolutely  continuous spectrum\index{continuous spectrum} of the Maxwell
equations.

However, this radiation
cannot last forever, since
it
irrevocably
carries the energy to infinity while
the total energy is finite. Therefore, in the long-time
limit only $ \om_ {nn '}=0 $ survives, which means exactly 
that we have
 single-frequency asymptotics (\ref{BS})
in view of  (\ref{psic}).

\br\label{rBB}
{\rm
Of course, these perturbation arguments cannot provide a rigorous justification
of the long-time asymptotics (\ref {BS}) for
the coupled Maxwell--Schr\"odinger equations.
In \ci{KK2006}--\ci{K2016}, we have justified similar
 single-frequency asymptotics
 for a list of 
 model nonlinear PDEs with the symmetry group $U(1)$.
Nevertheless, for the coupled Maxwell--Schr\"odinger equation 
such a justification is still an open problem.

}
\er

\subsection{Bohr' postulates as global attraction}

The perturbation arguments above suggest that the Bohr postulates can be treated as
 the long-time asymptotics 
\be \la{ate1}
(\bA (x, t), \psi (x, t)) \sim (\bA_ \pm (x), 
e ^ {- i \om_ \pm t} \psi_ \pm (x)), \qquad t \to \pm \infty
\ee
{\it for all finite-energy solutions} of the  Maxwell--Schr\"odinger equations
(\ref{MS}) in the case of {\bf static external potentials} (\ref{stat}). We conjecture that these asymptotics 
hold 
in  $H^1$-norms on every bounded region  of $\R^3$.

\br
{\rm
Experiments show that the transition time of quantum jumps (\ref {B1}) is of the order $10^{-8}s$, although the asymptotics  (\ref {ate1}) require infinite time.
We suppose that this discrepancy can be explained by the following arguments:
 \smallskip\\
 i) $10^{-8}$s is the transition  time between very small neighborhoods of
initial and final states, and
 \smallskip\\
 ii) during this time the atom emits an overwhelming part of the radiated energy.
 }
\er

The
asymptotics  (\ref{ate1}) have not been proved
 for the Maxwell--Schr\"odinger system
(\ref{MS}). On the other hand, similar asymptotics
are now proved for a number of model Hamiltonian nonlinear PDEs with the symmetry group $U(1)$. In next section we state a general conjecture which 
reduces to the asymptotics (\ref{ate1}) in the case of the Maxwell--Schr\"odinger system.

\bd
{\rm
{\it Stationary orbits} of the Maxwell--Schr\"odinger nonlinear system
(\ref {MS}) are
finite energy
solutions of  the form 
\be\la{Sorb}
(A (x), e ^ {- i \om t} \psi (x)) .
\ee
}
\ed
The existence of stationary orbits for the system (\ref {MS})
 was proved in \ci {CG2004} in the case of external potentials
\be \la {AAA}
\bA_\ext (x, t) \equiv 0, \qquad A _ {\rm ext} ^ 0 (x, t) = - \fr {eZ} {| x |}, \qquad \int | \psi_ \pm (x) | ^ 2 dx \le Z.
\ee

The asymptotics (\ref{ate1}) 
 mean that a {\it global attraction} to
the set of stationary orbits occurs. We suggest 
that a 
similar attraction takes place for
Maxwell--Dirac, Maxwell--Yang--Mills and other
coupled equations.
In other words, we suggest to interpret quantum stationary
states as the points and trajectories that constitute the {\it global attractor} of the corresponding quantum dynamical equations.

\subsection{The Einstein--Ehrenfest paradox and bifurcation of attractors}
An instant
orientation of  the atomic magnetic moment
during $ \sim 10 ^ {- 4} s $, when turning on the magnetic field
in the Stern--Gerlach experiments, led to
 a discussion in the `Old Quantum Mechanics',
because the classical model gave relaxation time
$ \sim 10 ^ 9 s $, taking into account the moment of inertia of the atom \ci {EE1922}.
In the  linear  Schr\"odinger's theory, this phenomenon also did not find a satisfactory explanation.
\smallskip

On the other hand, this instantaneous orientation is exactly in  line
 with  asymptotics (\ref {ate1}) for solutions of the
coupled nonlinear Maxwell--Pauli equations, i.e., the   Maxwell--Schr\"odinger  system with spin.
In the absence of the magnetic field,
the 
minimal eigenvalue\index{eigenvalue} 
is of  multiplicity 2, which implies 
that
 the 
manifold of the corresponding ground states\index{ground state} $C_1\psi_1^0+C_2\psi_2^0$  
with $|C_1|^2+|C_2|^2=1$ is of  dimension 3.
Let us stress that 
in this case the
 special role of the states 
with a fixed spin momenta $s=\pm \fr 12$ 
  is {\bf illusory}, since the eigenfuctions depend on the choice of the coordinates in the space $\R^3$.

 However, in a small magnetic field\index{magnetic field},
this minimal eigenvalue\index{eigenvalue} bifurcates\index{bifurcation of eigenvalue} 
into two simple eigenvalues. This suggests that 
 for the corresponding nonlinear Maxwell--Pauli equations
 an {\bf instant 
 bifurcation} of the global attractor\index{global attractor} 
 (i.e., of the set of  stationary orbits\index{stationary orbit} (\ref{Sorb}))
 occurs.
 In particular, {\bf new stationary orbits}, corresponding to
  the spin momentum $s=\pm \fr 12$, appear.

 One can expect that the Einstein--Ehrenfest paradox\index{Einstein--Ehrenfest paradox}
 can be explained by
 a similar bifurcation\index{bifurcation} of the attractor of
 the coupled nonlinear Maxwell--Pauli system\index{Maxwell--Pauli equations}
 describing the   Stern--Gerlach experiment\index{Stern--Gerlach experiment}.
Indeed,
when the magnetic field\index{magnetic field}
 is turned on, the {\bf structure of the attractor\index{attractor}
instantly changes}. Then the trajectory is
attracted to a suitable component of new attractor\index{attractor}
 with a certain spin value 
 that  
 corresponds to the `alternative~A' in the terminology of Einstein--Ehrenfest\index{Einstein--Ehrenfest paradox} \ci {EE1922}:
{\it ``... atoms can never fall into the state in which they are quantized {\bf not fully}".}
 This bifurcation is not related to any
rotation of electrons with
 moment of inertia\index{moment of inertia}, which
 explains the  Einstein--Ehrenfest paradox\index{Einstein--Ehrenfest paradox}.

\subsection{Attractors of dissipative and Hamiltonian PDEs }

Such an interpretation of the Bohr transitions as a global attraction is rather natural.
On the other hand, the existing theory of attractors of {\it dissipative systems}
 \ci{L1944}--\ci{T1997} does not help  in this case since
all funda\-men\-tal equations of quantum theory are {\it Hamiltonian}.
The global attraction for dissipative systems is caused by
energy dissipation. However, such a~dissipation
in the Hamiltonian systems is absent.

This is why the author has developed (in the years 1990--2020), 
together with his collaborators,
a novel theory
of global attractors for    {\it Hamiltonian}  PDEs, especially for the application
to  Quantum Theory problems.
All obtained results \ci{K1995a}--\ci{C2013}   for
the Hamiltonian equations rely on  
a thorough analysis of
energy radiation, 
 which
irrevocably carries the energy  to infinity and plays the role of  energy dissipation.
 A brief
survey of these results can be found in Section \ref{sattr}, and  detailed survey in \ci{K2016,KK2020}.

The results  obtained so far indicate
an explicit {\it correspondence} between the type of 
 long-time asymptotics of finite energy solutions and  the
symmetry group  of the equations.
We formalize this correspondence in our general conjecture (\ref{at10}).

\section{Conjecture on attractors of $G$-invariant equations}
Let us consider general 
 $G$-invariant  {\it autonomous} Hamiltonian nonlinear PDEs in $\R^n$ 
 of type
 \be\la{dyn}
\dot\Psi(t)=F(\Psi(t)),\qquad t\in\R,
\ee
with a Lie  symmetry group $ G $ acting
on a suitable Hilbert or Banach phase space $\cE$ via a linear 
representation $T$.
The Hamiltonian structure \index{Hamiltonian structure} means that
\be\la{Hamstr}
F(\Psi)=J D \cH(\Psi),\qquad J^*=-J,
\ee
where $\cH$ denotes the corresponding Hamiltonian functional.
The
$G$-invariance\index{G-invariance} means that 
\be\la{Ginv}
 F (T(g) \Psi) = T(g)F (\Psi),\qquad
 \Psi\in \cE
\ee
 for all $ g \in G $.
In that case, for any solution $ \Psi (t) $ of
equation (\ref {dyn}) the trajectory  $ T(g) \Psi (t) $ is also a solution,
so the representation commutes with the dynamical group 
$U(t):\Psi(0)\mapsto \Psi(t)$,
\be\la{comTU}
T(g)U(t)=U(t)T(g).
\ee
Let us note that
the theory of elementary particles deals systematically with the
symmetry groups
$ SU (2) $, $ SU (3) $, $ SU (5) $, $ SO (10) $ and others,
as well as with the group
$$
 SU (4) \times SU (2) \times SU (2),
 $$
  which is 
the symmetry group\index{symmetry group} of
`Grand Unification'\index{Grand Unification}, see \ci {HM1984}.
\smallskip\\
{\bf Conjecture  A.} (On attractors) {\it
For `generic' $G$-invariant autonomous equations (\ref{dyn}),
any finite energy solution $ \Psi (t) $  admits a long-time asymptotics
\begin {equation} \label {at10}
\Psi (t) \sim e ^ {\hat \lam_ \pm t} \Psi_ \pm, \qquad t \to \pm \infty,
\end {equation}
in the appropriate topology of the phase space $\cE$.
Here 
$\hat \lam_\pm=T'(e)\lam_\pm$, where 
$\lam_\pm$ belong to the corresponding 
 Lie\index{Lie algebra} algebra
$\mathfrak{g}$, 
while the $\Psi_\pm $ are some
{\bf limiting amplitudes\index{limiting amplitude}} depending on the  trajectory 
$\Psi(t)$ considered.
}
\smallskip

 In other words, all
 solutions of the type
$e ^ {\hat\lam t} \Psi$ with $ \lam\in \mathfrak {g} $
form a global attractor\index{global attractor} for {\bf generic} $G$-invariant Hamiltonian nonlinear
 PDEs\index{G-invariant equation} of type \eqref{dyn}.
This conjecture\index{conjecture on attractors} suggests
to define {\it stationary  $G$-orbits}\index{stationary  G-orbit} for equation \eqref{dyn} as solutions of the type
\begin {equation} \label {at10a}
\Psi (t) = e ^ {\hat \lam t} \Psi, \qquad t \in \R,
\end {equation}
 where $\lam$
belongs to the corresponding Lie algebra\index{Lie algebra}
$ \mathfrak {g} $.
 This definition leads to the corresponding
 {\it nonlinear eigenvalue problem}\index{nonlinear eigenvalue problem}
\begin {equation} \label {at10b}
\hat \lam \Psi = F (\Psi).
\end {equation}
In particular, for  the linear Schr\"odinger equation
with the symmetry  group  $U (1) $,
stationary orbits are solutions of the form $ e ^ {i \om t} \psi (x) $,
where $ \om \in \R $ is an eigenvalue of the Schr\"odinger operator, and $ \psi (x) $ is the corresponding eigenfunction. 
In the case of the symmetry group $ G = SU (3) $, the generator
(`eigenvalue') $ \lam $ is
$ 3 \times 3 $ -matrix, and solutions (\ref {at10a}) can be quasiperiodic in time.

  Note that
the conjecture  (\ref{at10}) fails for linear equations, i.e., linear  equations are exceptional,
 not `generic'!

\medskip

{\bf Empirical evidence.}
Conjecture (\ref{at10}) agrees with the Gell-Mann--Ne'eman theory of baryons  \cite{GM1962, Ne1962}.
 Indeed, in 1961
Gell-Mann and Ne'eman  suggested  using the symmetry group $SU(3)$ 
for the strong interaction of baryons  relying on the discovered parallelism between empirical data
for the baryons,
and the ``Dynkin scheme''
of Lie algebra $\mathfrak{g}=su(3)$ with $8$ generators (the famous `eightfold way').
This theory resulted in
the scheme of quarks in
quantum chromodynamics \cite{HM1984},
and in the prediction of a new baryon
with prescribed values of
its mass and decay products. This particle (the $\Omega^-$-hyperon)
was promptly discovered experimentally \cite{omega-1964}.

On the other hand, the elementary particles seem to describe long-time asymptotics
of quantum fields. Hence
the empirical correspondence between elementary
particles and generators of the Lie algebras
presumably gives an
evidence in favour of our general conjecture
(\ref{at10}) for equations with  Lie symmetry groups.

\section {Results on global attractors for nonlinear Hamiltonian PDEs}
\la{sattr}

Here we give a  brief survey of
rigorous results \ci{K1995a} - \ci{C2013}
obtained since 1990 that confirm conjecture  (\ref{at10})
for a list of model equations of type (\ref{dyn}).
The details can be found in \ci {K2016,KK2020}.

\begin{figure}[htbp]
		\begin{center}
			\includegraphics[width=0.7\columnwidth]{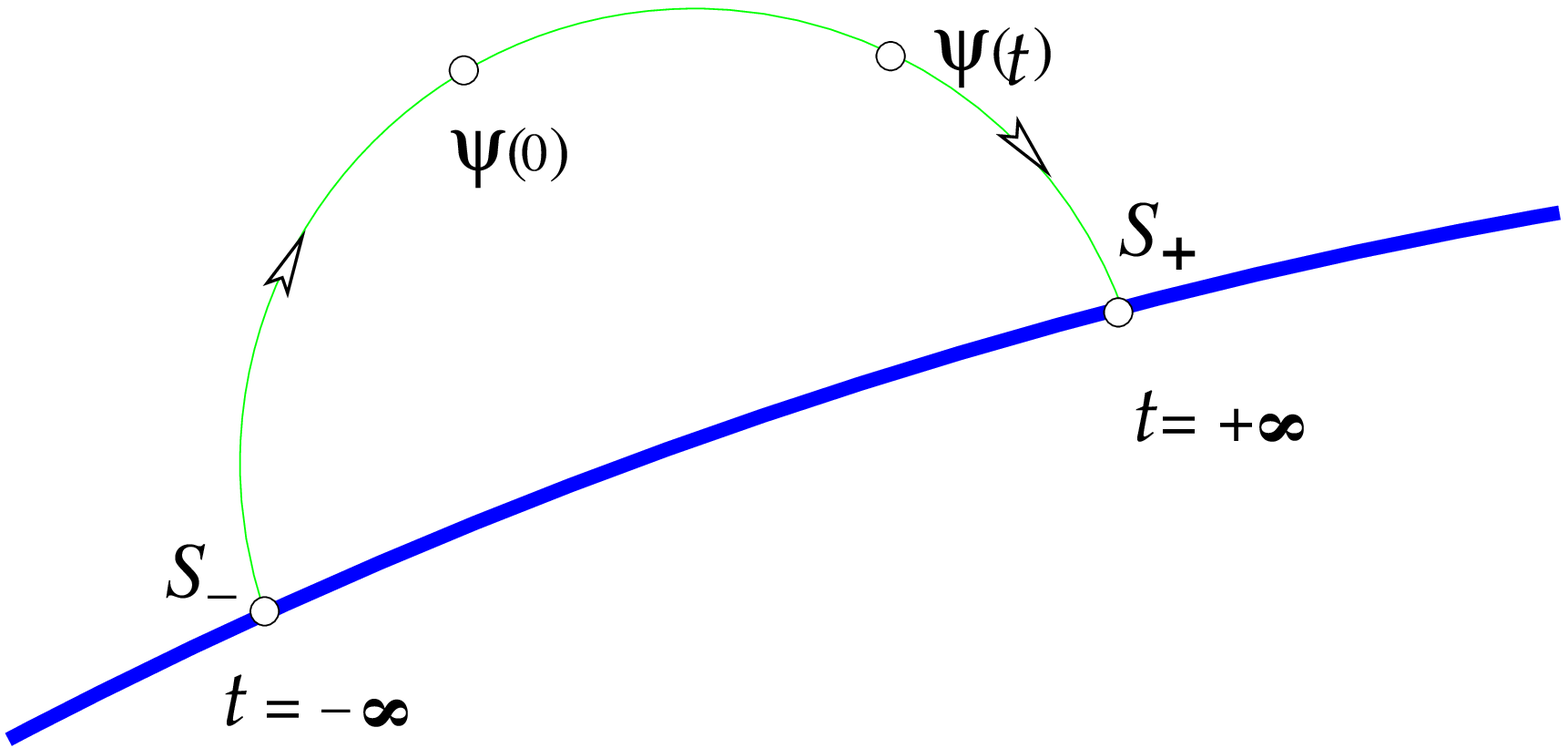}
			\caption{Convergence to stationary states.}
			\label{fig-1}
		\end{center}
	\end{figure}

 The results 
confirm the existence of finite-dimensional attractors
in the Hilbert or Banach phase spaces, and demonstrate
an explicit correspondence between
the long-time asymptotics and  the symmetry group $ G $ of equations.
\smallskip
The results obtained so far 
concern equations (\ref{dyn}) with
the following
four basic groups of symmetry:
the trivial symmetry group
$ G = \{e \} $, the translation group $ G = \R ^ n $ for 
translation-invariant equations, the unitary group $ G = U (1) $ for phase-invariant equations, and the orthogonal group $SO(3)$ for `isotropic' equations.
In these cases, the asymptotics (\ref{at10}) reads as follows.
 \smallskip\\
{\bf I. Equations with trivial symmetry group $G=\{e\}$.}
For such {\it generic equations}
the conjecture (\ref{at10})  means {\it global attraction to  stationary states}
 \begin {equation} \label {ate}
\psi (x, t) \to S_ \pm (x), \qquad t \to \pm \infty,
\end {equation}
as illustrated in Fig.~\ref {fig-1}.

\vspace{-30mm}

Here  the states $S_\pm=S_\pm(x) $  depend on  the trajectory $\Psi(t)$ under
consideration,
and the convergence holds in local 
seminorms\index{local seminorm} of type  $L^2(|x|<R)$ with any $R>0$.
This convergence cannot hold
in global norms\index{global norm} (i.e., in norms corresponding to $R=\infty$)  due to  
energy conservation\index{conservation of energy}.
The asymptotics\index{asymptotics} (\ref{ate}) can be symbolically written as the transitions\index{transitions} 
\be\la{Bohr}
S_-\mapsto S_+,
\ee
which
can be considered as the mathematical model of the Bohr\index{Bohr} `quantum jumps'\index{quantum jump} (\ref{B1}).

Such an attraction
was proved for a variety of model equations
in \ci {K1991}--\ci {KK2020c}: i) for a string coupled to nonlinear oscillators, ii) for a~three-dimensional wave equation coupled
to a charged particle and for the Maxwell--Lorentz equations, and also
iii) for
wave equation, and Dirac and Klein--Gordon equations
with concentrated nonlinearities. All proofs rely on the analysis of radiation\index{energy radiation} 
which  irreversibly carries energy
to infinity. The  details can be found  in the survey \ci{KK2020}. 

In all the problems considered,
the convergence (\ref{ate}) implies, by the Fatou theorem\index{Fatou theorem}, the 
inequality
\be\la{1.6''}
{\cal H}(S_\pm)\leq {\cal H}(Y(t))\equiv\const,\,\,\,t\in\R,
\ee
where $\cH$ is the corresponding Hamiltonian (energy) 
functional\index{Hamiltonian}. This inequality is an analog
 of the well known property of weak convergence\index{weak convergence} in 
Hilbert\index{Hilbert space} and Banach\index{Banach space} spaces. 
Simple examples show that
the strict inequality in (\ref{1.6''}) is possible, which
 means that an irreversible
  scattering of energy\index{energy radiation}  to infinity occurs.

 \bex
{\rm {\bf The d'Alembert waves\index{d'Alembert wave}.}
In particular, the asymptotics\index{asymptotics} (\ref{ate})
with the strict inequality (\ref{1.6''}) 
can 
easily be demonstrated for the d'Alembert\index{d'Alembert equation} equation 
$\ddot\psi(x,t)=\psi''(x,t)$
with general solution
\be\la{dalw}
\psi(x,t)=f(x-t)+g(x+t).
\ee
 Indeed, the  convergence 
$\psi(\cdot,t)\to 0$ in $L^2_{\rm loc}(\R)$ obviously
holds for all $f,g\in L^2(\R)$. On the other hand, 
 the convergence to zero in {\it global norms}\index{global norm}
obviously fails if $f(x)\not\equiv 0$ or $g(x)\not\equiv 0$. 
}
\eex

\bex {\bf Nonlinear Huygens Principle.}
{\rm
Consider solutions of  3D wave equation  with a unit propagation velocity and initial data with support in a ball $ | x | <R $. The corresponding
solution is concentrated in spherical layers $ | t | -R <| x | <| t | + R $. Therefore, the solution converges everywhere to zero as
$ t \to \pm \infty $, although its energy remains constant.
This convergence to zero is  known as the
{\it  strong  Huygens principle}. Thus, global attraction to stationary states
(\ref {ate}) is a generalization of the strong Huygens principle to nonlinear equations.
The difference is that for the linear wave equation, the limit  
 is always zero, while for nonlinear equations
the limit can be any stationary solution.
}
\eex

\br
{\rm
The proofs
in \ci {KSK1997} and \ci {KS2000}
rely on the relaxation of acceleration
\be \la {rel}
\ddot q (t) \to 0, \qquad t \to \pm \infty,
\ee
which has been known for about 100 years as `radiation damping'
in Classical Electrodynamics, but was first proved in \ci {KSK1997} and \ci {KS2000} for charged relativistic particle in a scalar field
and in the Maxwell field under the {\it Wiener Condition} on the particle charge density.
This condition is an
analogue of the `Fermi Golden Rule', first introduced by Sigal in the context of nonlinear wave- and Schr\"odinger equations \ci {Sigal}.
The proof of the relaxation (\ref{rel}) relies
on a novel application of the Wiener Tauberian theorem.
}
\er
\subsection{Group of translations $G=\R^n$}

For  {\it generic translation-invariant equations}
the conjecture (\ref{at10}) means
the {\it global attraction to solitons}
\begin {equation} \label {att}
\psi (x, t) \sim \psi_ \pm (x-v_ \pm t), \qquad t \to \pm \infty,
\end {equation}
where the
convergence holds in local seminorms
of type $ L ^ 2 (| x-v_ \pm t | <R) $ with any $ R> 0 $, i.e.,
{\it in the comoving frame of reference}.
A trivial example is provided by  the d'Alembert equation
$ \ddot \psi (x, t) = \psi '' (x, t) $ with 
general solution 
 (\ref{dalw}) corresponding to the asymptotics 
(\ref{att}) with $v_+=\pm 1$ and $v_-=\pm 1$.
\smallskip 

Such soliton asymptotics 
was first proved   for {\it integrable equations} (Korteweg--de Vries equation (KdV), etc), see \ci {EvH,Lamb80}.
 Moreover,
for  the Korteweg--de Vries equation 
more accurate soliton asymptotics in {\it global norms}
with several solitons were first discovered
 by Kruskal and Zabuzhsky in 1965 by
numerical simulation:
it is the decay to solitons
\be \la {attN}
\psi (x, t) \sim \sum_k \psi_ \pm (x-v ^ k_ \pm t) + w_ \pm (x, t), \qquad t \to \pm \infty,
\ee
where
$ w_ \pm $ are some dispersion waves.
\smallskip 

  Later on, such asymptotics
were proved by the method of {\it inverse scattering problem}
for nonlinear
\textbf {integrable} Hamiltonian
translation-invariant equations (KdV, etc.)
in the works of Ablowitz, Segur, Eckhaus, van Harten and others \cite {EvH,Lamb80}.
\smallskip

 For {\it non-integrable equations},
 the global attraction (\ref{att}) was established 
for the first time in
\ci{KS1998}--\ci{IKS2004a}
for three-dimensional wave equation coupled
to a charged particle and for the Maxwell--Lorentz equations.
The proofs
in \ci {KS1998} and \ci {IKM2004}
rely on  variational properties of solitons and their orbital stability, as well as on the  relaxation of the acceleration (\ref {rel})
under the Wiener
condition on the particle charge density.

 The multi-soliton asymptotics 
(\ref{attN}) for {\it non-integrable equations}
 were observed numerically 
 in  \ci{KMV2004}
for 1D {\it relativistically-invariant} nonlinear wave equations.

\subsection{Unitary symmetry group $G=U(1)$}

For {\it generic $U(1)$-invariant equations}
the conjecture (\ref{at10}) means
the {\it global attraction
to `stationary orbits'}
\begin{equation}\label{atU}
	\psi(x,t)\sim\psi_\pm(x) e^{-i \omega_\pm t} , \qquad t \to \pm \infty,
\end{equation}
where
 $\om_\pm\in\R$.
 Such  asymptotics are similar to 
  the Bohr transitions  between stationary orbits  (\ref{ate1}) of the coupled Maxwell--Schr\"odinger equations.

\begin{figure}[htbp]
	\begin{center}
		\includegraphics[width=0.9\columnwidth]{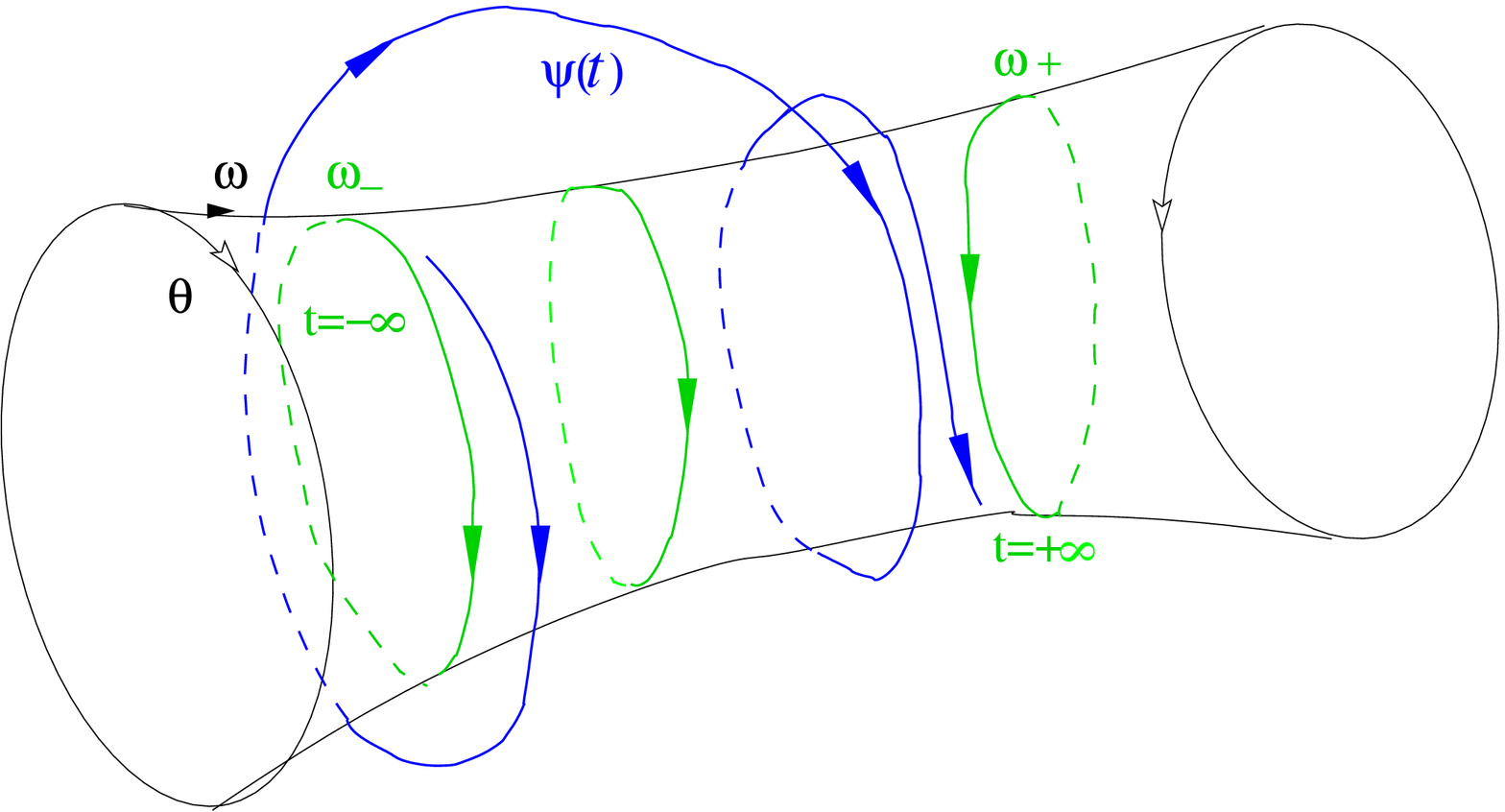}
		\caption{Convergence to stationary orbits.}
		\label{fig-3}
	\end{center}
\end{figure}

This asymptotics means that the
global attraction to the solitary manifold\index{global attraction to solitary manifold} formed by all {\bf stationary orbits\index{stationary orbit}} (\ref{SSO}) occurs.
The asymptotics\index{asymptotics} are considered
in the local seminorms\index{local seminorm} $L^2(|x|<R)$ with any $R>0$.
The global attractor\index{global attractor} is a smooth manifold formed by the circles
that are the orbits of the action of the 
symmetry group\index{unitary symmetry group} $U(1)$ as is illustrated in
 Fig.~\ref{fig-3}.

Such an attraction 
  was proved 
  for the first time
  i) in
\ci{K2003}--\ci{KK2010b} for the 
Klein--Gordon\index{Klein--Gordon equation}
  and Dirac\index{Dirac equation} equations coupled to
$U(1)$-invariant nonlinear oscillator\index{nonlinear oscillator}, 
 ii) in
\ci{C2013}, for discrete  approximations
of such coupled systems, i.e., for the corresponding difference schemes\index{difference scheme},
and iii) in \ci{K2017}--\ci{KK2020c}
for the wave, Klein--Gordon\index{Klein--Gordon equation}, and Dirac\index{Dirac equation} equations with concentrated nonlinearities\index{concentrated nonlinearity}.
More precisely, we have proved  
global attraction\index{global attraction to stationary orbits} to the 
{\it solitary manifold}\index{solitary manifold} of all 
stationary orbits\index{stationary orbit}, though
global attraction\index{attraction} to a particular  
stationary orbits\index{stationary orbit}, with fixed $\omega_\pm$, is still an open problem.

All
these results were proved under the assumption that the equations
are `strictly nonlinear'\index{strictly nonlinear equation}. For linear equations, the global attraction\index{global attraction to stationary orbits} obviously fails
if the discrete spectrum\index{discrete spectrum} consists of at least two different eigenvalues\index{eigenvalue}.

The proofs of all these results  rely on i) a nonlinear analog of the Kato theorem\index{nonlinear Kato theorem} on the absence of embedded eigenvalues\index{embedded eigenvalue}, ii) a new theory of multiplicators\index{multiplicator}
in the space of quasimeasures\index{quasimeasure} and 
iii) a novel application of the
Titchmarsh convolution theorem\index{Titchmarsh convolution theorem}.
\smallskip

\subsection{Orthogonal group $G=SO(3)$} 
In this case  (\ref{at10}) means  the long-time asymptotics\index{long-time asymptotics}
\begin{equation}\label{atSO3}
  \Psi(t)\sim  e^{-i \hat\Om_\pm t}\Psi_\pm , \qquad t \to \pm \infty,
\end{equation}
where $\hat\Om_\pm$ are 
suitable representations of 
real skew-symmetric
$3\times 3$ matrices  $\Om_\pm\in \mathfrak{so}(3)$.
This  means that
global attraction to `stationary $SO(3)$-orbits'\index{global attraction to `stationary $SO(3)$-orbits'} occurs.  
Such asymptotics are proved in \ci{IKS2004b} for the 
Maxwell--Lorentz equations with rotating 
particle\index{Maxwell--Lorentz equations with spinning particle}.
 \smallskip\\
{\bf Generic equations.}
We must still    specify the 
meaning of the 
term {\bf generic} in  our conjecture\index{conjecture on attractors} (\ref {at10})
 In fact, this conjecture
means that the asymptotics (\ref {at10}) hold for all solutions for
an {\it open dense set} of $G$-invariant equations\index{G-invariant equation}.
\smallskip \\
i) In particular, the asymptotics
 (\ref {ate}), (\ref {att}), (\ref {atU}) and (\ref{atSO3})
hold under appropriate conditions, which define some `open dense subset' of
$G$-invariant equations\index{G-invariant equation} with the four types of symmetry group\index{symmetry group} $G$.
This asymptotic expression
may break down if these conditions fail\,---\,this corresponds to some `exceptional'
equations\index{exceptional equation}. For example, global attraction  (\ref {atU}) breaks down for the linear Schr\"odinger equations\index{Schr\"odinger equation} with at least two different eigenvalues.
Thus, linear equations are exceptional\index{exceptional equation},
 {\bf not  generic}!
\smallskip \\
ii) The general situation is the following. Let a Lie group\index{Lie group} $ G_1 $ be a (proper) subgroup\index{subgroup} of some larger Lie group\index{Lie group} $ G_2 $. Then
$ G_2 $-invariant equations
form an `exceptional subset' among all 
$ G_1 $-invariant equations, and
the corresponding asymptotics (\ref {at10}) may be completely different.
For example, the trivial group\index{trivial group} $ \{e \} $ is a proper 
subgroup in $ U (1) $ and in $ \R ^ n $, and the asymptotic expressions
(\ref {att}) and (\ref {atU}) may differ significantly from (\ref {ate}).

\section{Wave-particle duality}
In his PhD of 1923, de Broglie 
suggested 
to
identify the beam of particles with a harmonic wave:
 \begin {equation} \label {dB1}
 \!\!\!\!\!\!\!\!
\left.
 \!\!\!\!\!\!\!\!\ba{c}
\mbox{\it a beam of free particles with momentum $ k $ and energy $ E $ is described }
\\
\mbox{\it with the wave function}\,\,
 \psi (x, t) = Ce ^ {i (k x- \om t)},\,\,{where}\,\, k=p/\hbar,\,\,\om=E/\hbar
\ea\right|.
\end {equation}
This identification was suggested 
by L. de Broglie for {\it relativistic particles}
as a counterpart to the Einstein
corpuscular treatment of light as a 
 beam of photons.

The  {\it nonrelativistic version} of this duality 
 was  the key source for the Schr\"odinger Quantum Mechanics.
 In this section,
   we discuss a possible treatment of this nonrelativistic 
   wave-particle duality 
 in the framework of
 semiclassical Maxwell--Schr\"odinger equations
 for the following phenomena:
 i) reduction of wave packets,
   ii) the diffraction of electrons, iii) acceleration of electrons in 
   the electron gun.

\subsection{Reduction of wave packets}

We suggest an appearance
of the wave-particle duality
relying on a {\it generalization of the conjecture} (\ref{at10}) to the case of
 translation-invariant Maxwell--Schr\"odinger system \eqref{MS} {\it without external
potentials},
i.e.,
$
\bA_\ext(x,t)\equiv 0$, $A^0_\ext(x,t)\equiv 0.
$
 In this case, the Schr\"odinger equation of \eqref{MS} becomes
 \be\label{S023}
 i \hbar \dot \psi (x, t)=
\fr 1 {2m} [- i \hbar \na\!-\! \ds \frac ec \bA ( x, t)]^2 \psi (x, t) \!+\! e A ^ 0(x, t) \psi (x, t)
,\quad x \in \R ^ 3.
 \ee
Now
 the symmetry group
 of the system \eqref{MS}
 is $G=\R^3\times U(1)$, and our general conjecture (\ref{at10}) should be strengthened similarly to
 (\ref{attN}) as the long-time asymptotics of each finite energy solution
\be \label {SA}
\!\!\!
\bA (x, t) \!\sim\!
\displaystyle\! \sum\limits_ {k} \bA_ \pm ^ k (x\!-\!v ^ k_ \pm t) + \bA_ \pm (x, t), \,\,
\psi (x, t) \!\sim\!
\displaystyle \!\sum\limits_ {k} \psi_ \pm ^ k (x\!-\!v ^ k_ \pm t)
e ^ {i \Phi_ \pm ^ k (x, t)} \!+ \!\psi_ \pm (x, t)
\ee
as $t \to \pm\infty$, where $ \bA_ \pm (x, t)$ and $\psi_ \pm (x, t)$
 stand for the corresponding {\it dispersion waves}.
  These asymptotics are considered in {\bf global energy norms}, and
  suggest to treat the solitons\index{soliton}
\be\la{solel}
( \bA_ \pm ^ k (x\!-\!v ^ k_ \pm t),\psi_ \pm ^ k (x\!-\!v ^ k_ \pm t)
e ^ {i \Phi_ \pm ^ k (x, t)})
\ee
as electrons\index{electron}. These asymptotics
provisionally correspond to the {\it reduction {\rm (}or collapse{\rm )} of wave packets}.

\subsection{Diffraction of electrons}

The most striking and direct manifestation of particle-wave duality is given by diffraction
 of electrons observed by
C. Davisson and L. Germer in their experiments of 1924--1927s.
In these first experiments, the electron beam was scattered by a crystal of nickel, and the reflected beam was fixed on a
film. The resulting images are similar to X-ray scattering patterns
(`lauegrams'), first obtained in 1912 by the method of Laue.
These experiments were the main motivation
 for Born's introduction of the probabilistic
interpretation of the wave function.

\begin{figure}[htbp]
\begin{center}
\includegraphics[width=0.5\columnwidth]{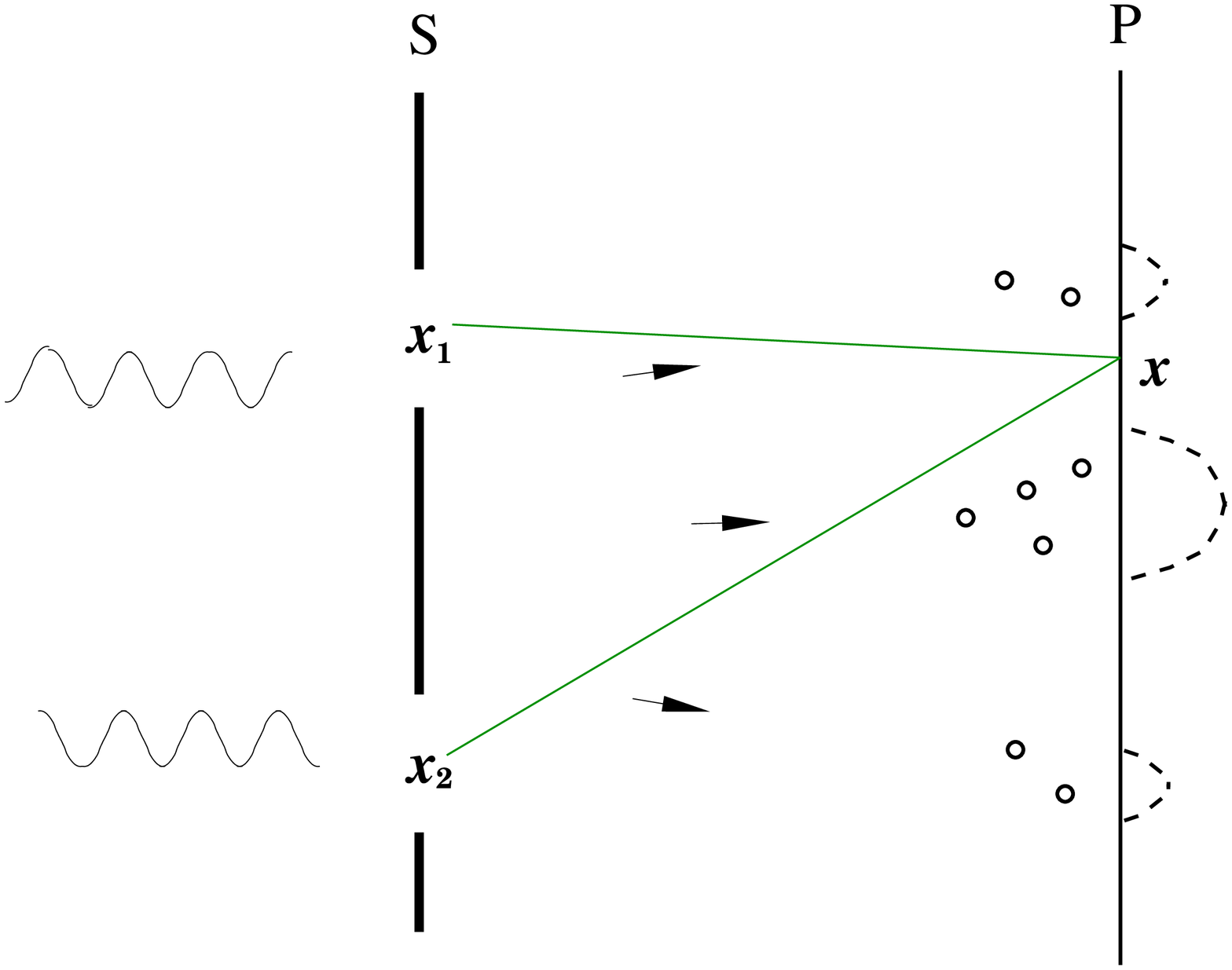}
\caption{Diffraction\index{diffraction} by double-slit.} \label{fT0}
\end{center}
\end{figure}

Later on, such experiments were also carried out with transmitted electron beams
scattered by thin gold and platinum crystalline films
(G.\,P.~Thomson, the 1937 Nobel Prize).
 Only recently R. Bach with collaborators
 first carried out  the two-slit diffraction of electrons \ci {BPLB2013} as is illustrated
by Fig. \ref{fT0}.

The traditional description of these experiments relies on the  following scheme:
\smallskip\\
I. The incident beam of electrons 
emitted by an electron gun
is described by the plane wave 
\be\label{inw}
\psi^{in}(x,t)=Ce^{i(kx-\om t)}
\ee
satisfying the free Schr\"odinger equation, 
with  wave number $k$ and frequency $\om$
given by the de Broglie relations (\ref{dB1}).
\smallskip\\
II. The diffraction of a plane wave in reflection 
by a crystal or in scattering by an aperture in the scattering screen.
\smallskip\\
III. The observation of the diffraction amplitude on the screen of registration.
\smallskip

We show in Section \ref{selg} below that 
the formation of the incident  plane wave 
(\ref{inw}) in the electron gun
can be explained by the quasiclassical asymptotics for the 
 {\it linear} Schr\"odinger equation.

Further, the diffraction in  step II is  traditionally described
by the {\it linear} Schr\"odinger equation, see \ci{ES1949}--\ci{H1988}. 
According to the {\it limiting amplitude principle}, the diffracted wave admits the asymptotics
\be \label {dif}
\psi_d ( x, t) \sim a_\infty ( x) e ^ {- i \om t}, \qquad t \to \infty,
\ee
where $ \om $ is the frequency of the incident wave (\ref{inw}). 
We calculated
 the diffraction amplitudes $ a_\infty ( x) $ 
 in the case of diffraction by the screen
using {\it the Kirchhoff approximation} (formulas (2.7.15) and (2.7.16)  of \ci{LQM-12}).
In particular,
for the two-slit diffraction,
 the corresponding  amplitude \ci[(2.7.20)]{LQM-12}
is in a fine {\it quantitative} agreement 
with the results of recent diffraction experiments
\ci{BPLB2013}.
Indeed, the maxima of
 $ |a_\infty ( x)| $ on
 the screen agree very well with that
of the diffraction pattern in
experiments \ci {BPLB2013}.

Thus, the arguments above rely on the linear Schr\"odinger theory.
On the other hand, the formation of the diffraction pattern in the step III 
is a {\it genuinely nonlinear effect} of the reduction of  wave packets
which
can be interpreted as
 the soliton asymptotics 
(\ref{SA}) for solutions of the 
coupled {\it nonlinear} Maxwell--Schr\"odinger equations.

\subsection{Quasiclassical asymptotics for electron gun}\la{selg}

Here we justify the
incident wave function (\ref{inw}) for electrons
emitted by an electron gun.
 The hot cathode emits electrons of small energy and after which the electrons are accelerated in an
electrostatic field $\bE(x)=-\na \Phi(x)$ in the region $\Om_1$
between the cathode and anode.
The electrons
cross the region $\Om_1$ and 
pass through an aperture in the anode
into the region $\Om_2$ behind the anode, where the
field $\bE(x)$ vanishes. In this region, the electrons interfere,
forming a diffraction pattern.
\smallskip

We will show that the diffraction amplitude
 coincides with the one corresponding
  to the free Schr\"odinger equation in the region $\Om_2$ with the incident wave (\ref{inw})
at the points of the aperture, where the
 corresponding
wave vector $k$ and frequency $\om$ are given by de Broglie's formulas (\ref{dB1}).
In this case, $E$ and $p$ denote the {\it kinetic energy and momentum} of the classical electron
 at the points of the aperture.

Namely, the
 electron wave function
 in the region $\Om_1\cup\Om_2$
 is a solution to the
corresponding Schr\"odinger equation
\be\label{S02}
 i \hbar \dot \psi (x, t)=
-\fr {\hbar^2} {2\cm} \De \psi (x, t) + e \Phi(x) \psi (x, t),
\quad x\in \Om_1\cup\Om_2,
 \ee
 where $\Phi(x)$ is the electrostatic potential vanishing on the cathode.
 This potential is constant in $\Om_2$, where the electrostatic field vanishes,
 so the equation (\ref{S02}) behind the anode reads
 \be\label{S02ba}
 i \hbar \dot \psi (x, t)=
-\fr {\hbar^2} {2\cm} \De \psi (x, t) + e \Phi_* \psi (x, t),\qquad x\in\Om_2,
 \ee
 where $\Phi_*$ is the value of the potential $\Phi(x)$ at the points of the aperture.
Further,
 we suppose that the solution admits the quasiclassical asymptotics
 \be\label{qca}
 \psi(x,t)\sim a(x,t)e^{i\fr{S(x,t)}\hbar},\qquad\hbar\to 0,
\ee
where $a(x,t)$ and $S(x,t)$ are slowly varying fuctions.
Substituting into (\ref{S02}),
we obtain in the limit $\hbar\to 0$
the corresponding Hamilton--Jacobi
equation for the phase function
\be\label{HJ2}
 -\pa_t S(x,t)=
\fr1{2\cm} [\na S (x,t)]^2 + e \Phi(x),\quad x\in\Om_1.
\ee
The solution is given by the action integral
over classical trajectories 
which satisfy
the Hamiltonian equations
\be\label{Heq}
\dot  x(t)=p(t)/\cm,\qquad \dot p(t)=-e\na\Phi( x(t)),
\ee
corresponding to the Hamiltonian function
\be\label{Hxp}
H( x,p)=\fr1{2\cm}p^2+e\Phi( x).
\ee
In particular, let us consider the trajectories
of emitted classical electrons
starting at time $t_0$ from all points $ x(t_0)$ of the cathode with momentum $ p(t_0)$
and passing
trough the aperture at the time
$t_*=t_0+T$. 
The
initial data
at time $t_0$
are related by $ p(t_0)=\na S( x(t_0))$ for all possible initial points $x(t_0)$. Hence
 we have for all trajectories,
\be\label{ker}
\na S( x(t),t)= p(t)
\ee
which follows from the Hamilton--Jacobi equation (\ref{HJ2}) (see \ci{A1989, Gol2001}).
The initial kinetic energy of the emitted electrons is small,
so we can assume that
\be\label{H00}
 H( x(t_0), p(t_0))=0.
\ee
Let us denote $ x_*:= x(t_*)$ and $ p_*:= p(t_*)$.
By the energy
conservation, we also have $H( x_*, p_*)=0$, and hence,
\be\label{HxpT}
\fr1{2\cm} p_*^2 =-e\Phi_*, \qquad \pa_tS( x_*,t_*)=0,
\ee
where the last identity follows from the Hamilton--Jacobi equation (\ref{HJ2}) with $x=x_*$ and $t=t_*$.
Now the Taylor expansion gives
\be\label{Tay}
S( x,t)\approx S( x_*,t_*)+ p_*( x- x_*),\qquad | x- x_*|+|t-t_*|\ll 1,
\ee
since $\pa_t S( x_*,t_*)=0$ by (\ref{HxpT}) and
\be\label{ker2}
\na S( x_*,t_*)= p_*
\ee
 by (\ref{ker}).
Therefore,
 {\it at the points of the  aperture} the wave function reads
\be\label{incw}
\psi( x,t)=a( x_*,t_*)e^{iS( x_*,t_*)/\hbar}e^{i p_*( x- x_*)/\hbar}\approx
Ce^{i p_*  x/\hbar},
\ee
since
$a( x_*,t_*)$ and $S( x_*,t_*)$ are
slowly varying functions.
\smallskip

Finally, the electrostatic potential $\Phi_*$
in the Schr\"odinger equation (\ref{S02ba})
behind the screen
can be eliminated
by the gauge transform
\be\label{gtr}
\psi( x,t)\mapsto \psi_*( x,t):=\psi( x,t)e^{ie\Phi_*t/\hbar}.
\ee
Indeed,  
the transformed function $\psi_*( x,t)$ 
behind the screen satisfies
the free Schr\"odinger equation
\be\label{S02baf}
 i \hbar \dot \psi_* ( x, t)=
-\fr {\hbar^2} {2\cm} \De \psi_* ( x, t),\qquad  x\in\Om_2.
 \ee
 The key observation is 
 that we have 
 the following asymptotics at the points of the small aperture
 \be\label{gtr2}
\psi_*( x,t):=\psi( x,t)e^{ie\Phi_*t/\hbar}\approx Ce^{i p_*  x/\hbar}e^{ie\Phi_*t/\hbar}
=Ce^{i( p_*  x-E_*t)/\hbar}
\ee
which follows from (\ref{gtr}) and (\ref{incw});
here
\be\la{EeP}
E_*=-e\Phi_*
\ee
 is the energy of the classical
electron at the aperture, and
$p_*$ is its momentum.
This justifies
the asymptotics of type (\ref{inw}) for $\psi( x,t)$
at the points of the  aperture
with 
\be\la{kp*}
k=p_*/\hbar,\qquad \om=E_*/\hbar
\ee
 which are exactly the
de Broglie relations (\ref{dB1}).
Finally,
the diffraction pattern corresponding to the charge densities $e|\psi( x,t)|^2$
and $e|\psi_*( x,t)|^2$
behind the aperture coincide by (\ref{gtr})  and the Born rule (\ref{probint}).

\br
{\rm
Note that the relations (\ref{EeP}), (\ref{kp*}) together with the first equation 
of (\ref{HxpT}) imply that 
\be\la{infr}
\hbar\om=\fr{\hbar^2k^2}{2\cm},
\ee
that is the incident wave (\ref{inw}) satisfies the free Schr\"odinger
equation.
}
\er

\section{Probabilistic interpretation}

In 1927, M. Born suggested the {\it Born rule} which is  probabilistic interpretation of the wave function:
\be\label{probint}
\left.\ba{c}
 \mbox{\it The probability of detecting an electron at a point $  x $ at time $ t $}\\
 \mbox{\it is proportional to $ | \psi ( x, t) | ^ 2 $.}
 \ea\right|
 \ee

\subsection{Diffraction current}
M.  Born proposed the probabilistic interpretation
  to describe the diffraction experiments
of C. Davisson and L. Germer of 1924--1927s.
Let us demonstrate that the rule (\ref{probint}) can be explained 
in the framework of these experiments
by
calculating the current (\ref{rj}) at the points of  the screen
of registration.

Indeed, the diffraction pattern is registered either by atoms of photo emulsion or by 
registration counters located on the screen. In all cases the {\bf rate of registration} 
 is proportional 
to the current density by definition. Thus, to explain the Born rule 
(\ref{probint}) in this situation, we must to show that the current density   (\ref {rj})
on the screen of registration
 is proportional to $|\psi(x,t)|^2$:
 \be\la{jpint}
 \bj (x,t)\sim |\psi(x,t)|^2(0,0,1)
 \ee
 if the screen lies in the plane $x_3=\const$.
The current density 
 (\ref {rj})
is approximately given by
\be \label {ja}
\bj (x,t) \approx\fr e \cm \psi(x, t) \cdot [- i \hbar \na \psi (x, t)].
\ee
Indeed, in formula (\ref {rj}), we have
$ \bA^\ext (x,t)= 0 $
since there is no external fields between the scatterer screen and the screen
of observation. The term with $ \bA (x,t) $ in (\ref {rj}) is  neglected, since its
contribution contains an additional small factor $\fr ec$.
\smallskip

For large times, the wave function is given by the limiting 
amplitude principle
 \be\label{reap12}
 \psi(x,t)\sim a_\infty (x)e^{-i\om t},\qquad t\to\infty.
 \ee
 The limiting amplitude $a_\infty ( x)$ near the diffraction screen 
is given by formula 
\be\label{Fraun}
a_\infty(x)\sim -\fr{ika_{in}(1+\cos\ov \chi) e^{ik|x|}} {(4\pi)^2|x|} \int_Qe^{-ik(\xi_1y_1+\xi_2 y_2)}dy,\qquad |x|\to\infty
\ee
similar to formula (28) of \ci[Section 8.3.3]{BW}
which describes the {\it Fraunhofer diffraction}.
Here
$(\xi_1,\xi_2,\xi_3):=x/|x|$
and  $\cos\ov\chi=\xi_3>0$; $Q$ is the aperture in the scattering plane $x_3=0$ and  the diffraction screen lies in the plane 
$x_3=\const\gg 1$.
This formula implies that for bounded $|(x_1,x_2)|\le C$
the limiting
amplitude  admits the following asymptotic representation, 
\be\label{Fraunas}
a_\infty(x)\sim \fr{b(\xi_1,\xi_2)}{|x|}e^{ikx_3},
\qquad x_3\to\infty.
\ee
Here the amplitude $b(\xi_1,\xi_2)$ is a slowly varying function of transversal variables
$(x_1,x_2)$ for large $x_3$, so asymptotically
\be\label{Fraunas2}
\na a_\infty(x)\sim ik a_\infty(x) (0,0,1)
\quad\mbox{\rm for } |(x_1,x_2)|\le C
\quad{\rm and} \quad x_3\to\infty.
\ee 
 Hence, (\ref{ja}) gives
\be \label {ja2}
\bj ( x,t) \approx \fr {e \hbar|k|} \cm  | a_\infty ( x) | ^ 2(0,0,1),\qquad t\to\infty
\ee
since the screen of registration is sufficiently far from the screen of scattering.
Finally, (\ref {reap12}) implies that for large times we have
 $ |a_ \infty (x)|^2\approx | \psi (x, t) | ^ 2 $, so
 (\ref{ja2}) implies (\ref{jpint}), which
 explains the Born rule\index{Born rule} (\ref{probint}).

\br
 {\rm
 Our arguments above justify 
 the Born rule (\ref{probint}) in the framework of 
 the diffraction experiments\index{diffraction experiment}
 with a plane screen in  the configuration of Fig. \ref{fT0}.
 For a curved screen, the same arguments suggest that 
 the rate of registration is proportional to the current and to $\cos\theta$,
 where $\al$ is
 the angle of incidence (the angle between the direction of the  electron beam and the normal to the screen).
  
}
 \er

\subsection{Discrete registration of electrons\index{registration of electrons}}
In 1948, the probabilistic interpretation\index{probabilistic interpretation} was given
new content and confirmation
by the experiments of L. Biberman\index{Biberman}, N. Sushkin\index{Sushkin} and V. Fabrikant\index{Fabrikant}
\ci {BSF1949} with  an electron beam\index{electron beam} of very low intensity.
Later, similar experiments were carried out by
R. Chambers\index{Chambers}, A. Tonomura\index{Tonomura}, S. Frabboni\index{Frabboni},
R. Bach\index{Bach} and others \ci{Chambers60, Tonomura89, FGP07, BPLB2013}.
In these experiments,
the diffraction pattern\index{diffraction pattern} is created as an average in time of random
discrete registration\index{discrete registration} of individual electrons.

We suggest below two possible treatments of
the probabilistic interpretation\index{probabilistic interpretation} in these 
experiments.
These treatments
rely respectively
\smallskip\\
i) on a random interaction with the counters\index{counter}, and 
\smallskip\\
ii) on
the {\bf soliton conjecture}\index{soliton conjecture} (\ref{SA})
in the  framework
of the coupled 
translation-invariant
Maxwell--Schr\"odinger equations\index{Maxwell--Schr\"odinger equations} \eqref{MS}.
\smallskip

However, the corresponding
rigorous justification is still an open problem.
\smallskip\\
{\bf I.  Interaction with counters.}
One possible  explanation of the discrete registration\index{discrete registration} is a
random triggering of a) registration counters located at the screen points,
or
b) atoms of the photo emulsion.
  In both cases,  the probability\index{probability} of triggering
  must be 
 proportional to the electric current\index{electric current} near the screen of observation. 
\smallskip\\
{\bf II. Reduction of wave packets.}
In the space
between the scatterer screen and the screen of observation, the {\bf external fields\index{external field} vanish}. 
Hence,  in this space, the Maxwell--Sch\"odinger system  is translation-invariant,
which is the case of our conjecture (\ref{SA})
on the decay to solitons--electrons
  (\ref{solel}),
see Fig.~\ref{fT0}.
Such a decay should be regarded as
a~random process, since it is subject to microscopic fluctuations.

 
\subsection{Superposition principle\index{superposition principle} as the linear approximation}
The treatment {\bf II} of the discrete registration  of electrons\index{diffraction of electrons} in the previous section
is  not self-consistent. Indeed,
the justification of the formula (\ref{ja2}) relies on the 
{\bf linear} Schr\"odinger equation\index{Schr\"odinger equation}, while the reference to 
the soliton asymptotics (\ref{SA}) involves 
 the {\bf nonlinear} Maxwell--Schr\"odinger equations\index{Maxwell--Schr\"odinger equations}.

We suggest the following  argument reconciling this formal contradiction: 
formula (\ref{ja2}) holds in the {\it linear approximation}, while
the soliton asymptotics (\ref{SA}) holds in the next approximation of the 
nonlinear Maxwell--Schr\"odinger equations.

\br
{\rm

The formula (\ref{ja2}) for the linear Schr\"odinger equation relies on the 
  {\bf superposition principle\index{superposition principle}}
which 
is the traditional argument
claiming that  quantum mechanics is {\bf absolutely linear}!
On the other hand, the 
discrete
registration of electrons
cannot be described by the linear Schr\"odinger equation\index{Schr\"odinger equation}.

}
\er

 \noindent {\bf Acknowledgements.}
The author thanks Sergey Kuksin, Alexander Shnirelman and Herbert Spohn for long-term discussions.

\bigskip\bigskip



\noindent{\bf\Large References}

\vspace{-3mm}


\bigskip

\noindent {\it Faculty of Mathematics, University of Vienna,  Oskar-Morgenstern-Platz 1, 1090 Wien, Austria.}
\vskip 0.3cm
\noindent {\it e-mail}: alexander.komech@univie.ac.at
\vskip 0.5cm

\end{document}